\documentclass[5p,12pt]{elsarticle}

\usepackage{amssymb}
\usepackage{amsmath}

\journal{Control Engineering Practice}

\begin{document}

\begin{frontmatter}

\title{Chance-constrained Stochastic MPC of Astlingen Urban Drainage Benchmark Network\tnoteref{t1}}
\tnotetext[t1]{This document is the results of the research project funded by the Spanish State Research Agency through the María de Maeztu Seal of Excellence to IRI (MDM-2016-0656), internal project of TWINs, and also supported by Innovation Fond Denmark through the Water Smart City project (project 5157-00009B).}

\author[1]{Jan Lorenz Svensen\corref{cor1}}
\ead{jlsv@dtu.dk}
\author[2]{Congcong Sun}
\ead{congcong@upc.edu}
\author[2,3]{Gabriela Cembrano}
\ead{gabriela.cembrano@upc.edu}
\author[2]{Vicen\c c Puig}
\ead{vicenc.puig@upc.edu}
\address[1]{Department of Applied Mathematics and Computer Science, Technical University of Denmark, Richard Petersens Plads 324, 2800 Kongens Lyngby, Denmark}

\address[2]{Advanced Control Systems Group, the Institut de Rob\`otica i Informàtica Industrial (CSIC-UPC), Llorens i Artigas, 4-6, 08028 Barcelona, Spain}

\address[3]{CETaqua, Water Technology Centre, Barcelona, 08904, Spain}

\cortext[cor1]{Corresponding Author}

\begin{abstract}
In urban drainage systems (UDS), a proven method for reducing the combined sewer overflow (CSO) pollution is real-time control (RTC) based on model predictive control (MPC). MPC methodologies for RTC of UDSs in the literature rely on the computation of the optimal control strategies based on deterministic rain forecast. However, in reality, uncertainties exist in rainfall forecasts which affect severely accuracy of computing the optimal control strategies. Under this context, this work aims to focus on the uncertainty associated with the rainfall forecasting and its effects. One option is to use stochastic information about the rain events in the controller; in the case of using MPC methods, the class called stochastic MPC is available, including several approaches such as the chance-constrained MPC method. In this study, we apply stochastic MPC to the UDS using the chance-constrained method. Moreover, we also compare the operational behavior of both the classical MPC with perfect forecast and the chance-constrained MPC based on different stochastic scenarios of the rain forecast. The application and comparison have been based on simulations using a SWMM model of the Astlingen urban drainage benchmark network.
\end{abstract}




\end{frontmatter}

\section{Introduction}
Regarding the state-of-the-art during the last couple of decades, Model Predictive Control (MPC) \cite{JM02} has been proved beneficial for the optimal operation of urban drainage systems (UDS)\cite{HFC}\nocite{JLS1,JLS2,GRJ,Sun17,Sun18,OMB,MPB}-\cite{CXJ}. Those studies use different types of modeling and optimization techniques to compute the best control actions, based on models and forecasts, which are subject to uncertainty. However, up to now, most of the MPC applications of UDS are based on deterministic rain forecasts without considering uncertainties, which may risk in introducing sub-optimal or undesired behaviors to the MPC solutions \cite{HFC}\nocite{JLS1,JLS2,GRJ,Sun17,Sun18,OMB,MPB,CXJ,Sun182,Sun172,Sun20,JDB17,CPJ}-\cite{Overloop6}.
 For a more realistic scenario, uncertainty has to be considered as a part of the UDS. The way how the uncertainty is treated by the control, becomes an important design decision: using a stochastic approach, or robustly operating on worst-case assumptions.

While the basic formulation of MPC is deterministic, 
 how to handle uncertainty in MPC has been researched for many years \cite{JLS4}-\nocite{Grosso2014,AGW06,DD06,ECK1,CGP09,Mesbah2016,KC16,Wan02,Magni03}\cite{ZSun18}. This has resulted in several different methods for handling uncertainty divided into two categories; the group of the methods known collectively as robust MPC\cite{KC16}-\nocite{Wan02,Magni03}\cite{ZSun18}, and the group of methods known as stochastic MPC\cite{JLS4}-\nocite{Grosso2014,AGW06,DD06,ECK1,CGP09,Mesbah2016}\cite{KC16}. The first group essentially considers the worst-case scenario and operates conservatively so that the solution is optimal for all possible realizations of the uncertainty. The second group addresses the uncertainty by using knowledge about the uncertainty, such as its distribution to only take the statistical likely scenarios into account for the control.

In this work, we will focus on a method from the group of stochastic methods known as chance-constrained MPC (CC-MPC)\cite{JLS4}-\nocite{Grosso2014,AGW06}\cite{DD06} to operate the UDS in order to reduce pollution to the receiving waters through minimization of the combined sewer overflows (CSO). Given that the CSOs are purely dependent on the volumes and flows of the system; the overflow constraints are intrinsically feasible and probabilistic insenstive, when CC-MPC is applied directly. We will therefore use the revised CC-MPC formulation\cite{JLS4} in this work.

In our previous work\cite{JLS2}, an MPC methodology was implemented and tested on a SWMM model of the Astlingen urban drainage benchmark network \cite{MMMU}, where the goal was to minimize the CSOs volume of the system, while maximizing the amount of treated wastewater by the wastewater treatment plant (WWTP). We obtained good results from this, in comparison with other real-time control strategies. In this paper, we return to the Astlingen urban drainage system for applying stochastic MPC using chance-constrained method regarding the uncertainty of rainfall forecast, and comparing the performance of CC-MPC with uncertain forecasts against the performance of the deterministic MPC with a perfect forecast. The key performance indexes considered are the CSO volume, and the volume received by the WWTP.

In this paper, the following mathematical notations are used. $\overline{f} $ indicates the maximum of a given function $f(x)$, $\beta$ represents the volume-flow coefficient\cite{singh1988}, and bold font is used to indicate vectors. The formulation $\|\textbf{x}\|_A^2=\textbf{x}^T A\textbf{x}$ is the weighted quadratic norm of \textit{x}. The superscript \textit{u} indicates control variables, superscript \textit{w} indicates CSO elements, and the superscripts \textit{in} and \textit{out} indicate inflow and outflow related flow, respectively. The letters \textit{V} and \textit{q} indicate variables of volume and flow respectively, while the variables written with \textit{w} are inflows from catchments. The notation $\Delta T$ and the subscript \textit{k} represent the sampling time of the system and the sample number respectively.

\section{Internal model of the Astlingen Benchmark Network}
\begin{figure}
\includegraphics[width=0.5\textwidth,trim={1.5cm 0.2cm 0.2cm 0cm},clip]{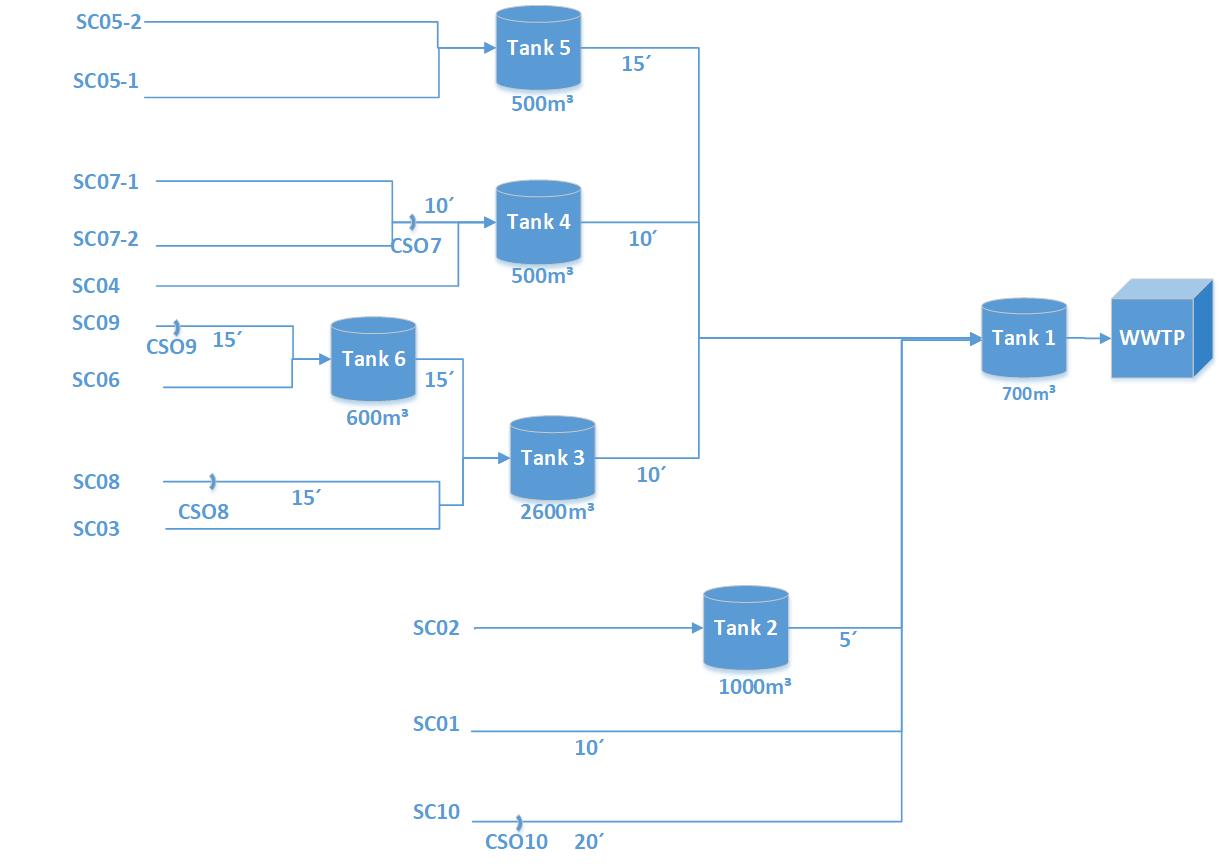}
\caption{A scheme of the Astlingen Benchmark Network\cite{MMMU} showing the interconnections between tanks, pipes and the WWTP, with CSOs coming from the six tanks and the four pipes noted CSO7 to CSO10. The delay between tanks and/or pipes are noted by x' in minutes.}\label{fig:1}
\end{figure}

The Astlingen urban drainage network consists 
 of six tanks and a single outflow towards a WWTP (see Figure \ref{fig:1}). In between and upstream of the tanks there 
are pipes of varying lengths, causing flow delays in the system. The system also consists of four pipes with CSO capabilities. The control variables of the system are the outflow of tanks 2, 3, 4, and 6. The desired operation of the system is to have the least amount of CSO as possible, and secondly having the largest 
amount of wastewater being sent to the WWTP. For designing an MPC controller for the system, an internal model describing the dynamics and constraints of this system is required, typically a simplified model of the system capturing the main dynamic behaviours is used.

From Figure \ref{fig:1}, it is clear that the system can be deduced to be uncontrollable (passive) in the sections upstream the tanks; therefore, the internal model will be limited to only covering the tanks of the system. The internal model is constructed with the same modular approach as used in previous works\cite{JLS2}. In the internal model, the CSO are treated as optimization variables through a penalty approach\cite{HFC}. The elements of the internal model consist of the following parts: linear reservoir tanks and pipes with delays that are described below.

In CC-MPC, the internal model of the deterministic MPC mentioned above is extended with a process equation of the variance of the dynamics, while the dynamics are replaced with the expectation of the dynamics. The constraints are reformulated either as the expectation of the constraint or as a probabilistic version of the constraints. The prior is in general used for equality constraints, while the latter is used for inequality constraints.

In this work, the run-off flows ($w$, covering runoff and passive flows) generated by forecasted rainfalls are the disturbance, involving uncertainty.  We will assume this uncertainty to follow a normal distribution, which is commonly used to interpret fluctuations in measured or forcasted variables\cite{KW2016,SP2020}. Then, for uncertainties following a normal distribution, the probabilistic constraints can be written deterministically as shown in (\ref{eq:1}), using the expectation $E\{x\}$ and standard deviation $\sigma\{x\}$ of the stochastic variable X, as well as the quantile function $\Phi^{-1}(x)$ of the standard normal distribution on the desired probability confidence level $\gamma$
\begin{equation}
\label{eq:1}Pr(X\leq x )\geq\gamma\Leftrightarrow x \geq E\{X\}+\sigma\{X\}\Phi^{-1}(\gamma)
\end{equation}
Furthermore, the only sources of uncertainty considered in the formulation of the internal model for the CC-MPC are the initial states of the system and the inflow from the run-off sources such as catchments. It is further assumed that the different sources of uncertainties are independently distributed, in both spatial and temporal sense.

\subsection{Linear Reservoir Tank - passive outflow}
The linear reservoir model has either a passive outflow or a controlled outflow and is based on mass-balance to describe the dynamics of tank volume. The volume of the tank $V_k$ is driven by the inflow $q_k^{in}$ and the weir overflow $q_k^w$. In the case of passive outflows, the outflow is controlled by gravity, and is assumed linear with a volume-flow coefficient\cite{singh1988} defined as  $\beta=\overline{q}^{out}/\overline{V}$. 

For the passive outflow case, the volume update and the outflow are defined by: 
\begin{align}
\label{eq:2} V_{k+1}=(1-\Delta T\beta) V_k+\Delta T(q_k^{in}-q_k^w ) \\
\label{eq:3} q_k^{out}=\beta V_k
\end{align}
The constraints of the reservoir are based on the physical constraints with the tank limits given by
\begin{align}
\label{eq:4} 0\leq (1-\Delta T\beta) V_k+\Delta T(q_k^{in}-q_k^w )\leq \overline{V} \\
\label{eq:5} 0\leq q_k^w
\end{align}

\subsubsection{CC-MPC formulation}
Utilizing the revised CC-MPC formulation\cite{JLS4} mentioned earlier, the passive reservoir model can be reformulated, such that the volume update and the outflow are defined by their expectation and variance given by 
\begin{align}
\label{eq:6}E\{V_{k+1} \}&=(1-\Delta T\beta)E\{V_k\}+\Delta T(E\{q_k^{in}\}-q_k^w )\\
\label{eq:7}E\{q_k^{out} \}&=\beta E\{V_k\}\\
\label{eq:8}\sigma^2\{V_{k+1}\}&=(1-\Delta T\beta)^2 \sigma^2 \{V_k \}+\Delta T^2 \sigma^2 \{q_k^{in} \}\\
\label{eq:9}\sigma^2\{q_k^{out}\}&=\beta^2 \sigma^2 \{V_k\}
\end{align}

The stochastic interpretation of the physical constraints is given by (\ref{eq:10})-(\ref{eq:14}), utilizing slack variables for guaranteeing feasibility\cite{JLS4}. 

The stochastic constraint for the lower limit of the tank is given by (\ref{eq:10}), while the upper limit is given by (\ref{eq:11}) and (\ref{eq:12}). The first one is a stochastic constraint for avoiding weir overflow $q_k^w$, while the latter is an expectation constraint defining the expected overflow
\begin{align}
\label{eq:10}\begin{split}\sigma\{(1-\Delta T\beta) V_k+\Delta Tq_k^{in} \} \Phi^{-1} &(\gamma)-s_k\leq\\ (1-\Delta T\beta)E\{V_k\}+\Delta T(E\{q_k^{in}\}&-q_k^w )
\end{split}\\
\label{eq:11}\begin{split}
(1-\Delta T\beta)E\{V_k\}+\Delta TE\{q_k^{in}\}\leq& \\ \overline{V}-\sigma\{(1-\Delta T\beta) V_k+\Delta Tq_k^{in} \}&\Phi^{-1} (\gamma)+c_k
\end{split}\\
\label{eq:12}(1-\Delta T\beta)E\{V_k \}+\Delta T(E\{q_k^{in} \}&-q_k^w )\leq\overline{V}\\
\label{eq:13}s_k\leq\sigma\{(1-\Delta T\beta) V_k+\Delta Tq_k^{in}& \} \Phi^{-1} (\gamma)\\
\label{eq:14}0\leq q_k^w,s_k,c_k
\end{align}
The limits on the slack variables $s_k$, $c_k$ are given by (\ref{eq:13}) and (\ref{eq:14}). For the control of the Astlingen model, Tank 1 and Tank 5 are considered tanks with passive outflow.

\subsection{Linear Reservoir Tank - Controlled outflow}
For a linear reservoir tank with controlled outflow, the volume is driven by the inflow $q_k^{in}$, the control flow $q_k^u$ and the weir overflow $q_k^w$. The volume update and outflow are defined by 
\begin{align}
\label{eq:15}V_{k+1}=V_k+\Delta T(q_k^{in}-q_k^u-q_k^w ) \\
\label{eq:16}q_k^{out}=q_k^u 
\end{align}
and the physical limits on the tanks and control are given by 
\begin{equation}
\label{eq:17}0\leq V_k+\Delta T(q_k^{in}-q_k^u-q_k^w )\leq \overline{V}
\end{equation}

The limits of the control including two upper limits of the control flow are defined as
\begin{align}
\label{eq:18}0\leq q_k^u\leq \overline{q}^u\\
\label{eq:19}q_k^u\leq \beta V_k \\
\label{eq:20}0\leq q_k^w
\end{align}
where the first one establishes the physical limit of the outflow pipe, and the other one a linear Bernoulli expression given by the volume-flow coefficient $\beta$.

\subsubsection{CC-MPC formulation}
The controlled reservoir model can be formulated for CC-MPC as below, considering that the volume update and outflow are defined by the expectation and variance 
\begin{align}
\label{eq:21}E\{V_{k+1}\}&=E\{V_k\}+\Delta T(E\{q_k^{in}\}-q_k^u-q_k^w)\\
\label{eq:22}E\{q_k^{out}\}&=q_k^u\\
\label{eq:23}\sigma^2\{V_{k+1}\}&=\sigma^2\{V_k\}+\Delta T^2 \sigma^2 \{q_k^{in}\}\\
\label{eq:24}\sigma^2 \{q_k^{out}\}&=0
\end{align}
Note that the outflow variance is zero, due to the control.

According to the reformulation\cite{JLS4}, the stochastic version of the physical constraints is given by 
\begin{align}
\label{eq:25}0\leq E\{V_k\}+\Delta T(E\{q_k^{in}\}&-q_k^u-q_k^w )\\
\label{eq:26}\begin{split}
 E\{V_k\}+\Delta T(E\{q_k^{in}\}-q_k^u )\leq&\\ \overline{V}-\sigma\{V_k+\Delta Tq_k^{in} \}& \Phi^{-1)}(\gamma)+c_k
\end{split}\\
\label{eq:27}E\{V_k\}+\Delta T(E\{q_k^{in}\}-q_k^u&-q_k^w )\leq \overline{V} \\
\label{eq:28}0\leq q_k^u\leq q^u \\
\label{eq:29}q_k^u\leq\beta E\{V_k\}-\beta\sigma\{V_k\}& \Phi^{-1} (\gamma)+s_k\\
\label{eq:30}s_k\leq  \beta \sigma\{V_k\}& \Phi^{-1} (\gamma) \\
\label{eq:31}0\leq q_k^w,c_k, s_k
\end{align}
where the slack variables are limited by (\ref{eq:30}) and (\ref{eq:31}). The constraints (\ref{eq:25})-(\ref{eq:27}) define the upper and lower limits of the tank, in a similar way as (\ref{eq:10})-(\ref{eq:12}). The control limits are defined by (\ref{eq:28}) and (\ref{eq:29}).

\subsubsection{Decoupling of slack variables}
In (\ref{eq:25}), the lower limit of the tank is given as expectation constraint, while in (\ref{eq:10}) it was expressed in a probabilistic manner. The change is due to the interconnections of the slack variables of the upper and lower constraints as follows 
\begin{equation}
\label{eq:32} s_k\leq c_k+ \overline{V}-\Delta T q_k^w
\end{equation}
where the upper slack is forced to be active if the lower slack is too large.

This can lead to an undesired trade-off during optimization when the uncertainty term is too large. This can be solved by a rescaling of the optimization weights or by reformulating the probability constraint. The latter was used here. The probability of the tank volume being above zero (\ref{eq:33}) can be rewritten 
\begin{equation}
\label{eq:33}\begin{split}
Pr(0\leq V_k+\Delta T(q_k^{in}-q_k^u-q_k^w ))\\= Pr(\Delta Tq_k^u\leq V_k+\Delta T(q_k^{in}-q_k^w ))\geq\gamma
\end{split}
\end{equation}
by considering that the tank volume $V_k$ are always below the upper tank limit, given that any volume above it would have turned into an overflow. This leads to the volume only decreases, when the control flow is used, i.e.
\begin{equation}
\label{eq:34} V_k\leq V_k+\Delta T(q_k^{in}-q_k^w )
\end{equation}
From here, we can replace (\ref{eq:33}) with a stricter and simpler probability as follows
\begin{equation}
\label{eq:35}\begin{split}
 Pr(0\leq V_k+\Delta T(q_k^{in}-q_k^u-q_k^w ))\\\geq Pr(\Delta Tq_k^u\leq V_k )\geq \gamma
\end{split}
\end{equation}

By multiplying with the volume-flow coefficient $\beta$ and assuming that $\beta\Delta T\leq1$, the probability constraint can be rewritten even stricter. The assumption is fair, given that if the opposite is true, then the volume can become negative. The resulting probability constraint
\begin{equation}
\label{eq:36} Pr(\beta\Delta Tq_k^u\leq\beta V_k )\geq Pr(q_k^u\leq\beta V_k )\geq\gamma
\end{equation}
can be recognized as (\ref{eq:29}), the stochastic version of one of the upper control limits. This indicates that if (\ref{eq:29}) holds so does (\ref{eq:36}), and therefore (\ref{eq:33}) would be a duplicate. For this reason, (\ref{eq:33}) can be replaced with the expectation constraint given in (\ref{eq:25}), for the inclusion of the lower limit of the tank. 

\subsection{Pipe with delays}
In the Astlingen network \cite{MMMU}, the tanks and upstream catchments 
are connected through pipes. The presence of these pipes introduces delays in the flows to the tanks from the upstream parts of the system.
 The importance of these delays depend on the chosen sampling time. Delays $\eta$ of exactly one sampling can be described by 
\begin{align}
\label{eq:37} \eta_{k+1,i}  = q_{k,i}^{in} \\
\label{eq:38} q_{k,i}^{out}=\eta_{k,i}
\end{align} 
where delays of multiple sampling times, can be constructed as a cascade of single delays
\subsubsection{CC-MPC formulation}
For the CC-MPC, the delay equations are replaced by their expectations 
\begin{align}
\label{eq:39} E\{\eta_{k+1,i} \}  = E\{q_{k,i}^{in} \} 	\\
\label{eq:40} E\{q_{k,i}^{out} \}=E\{\eta_{k,i} \} 	
\end{align}
In addition, the variance of the delay equations are given by 
\begin{align}
\label{eq:41} \sigma^2 \{\eta_{k+1,i}\} = \sigma^2 \{q_{k,i}^{in}\}\\
\label{eq:42} \sigma^2 \{q_{k,i}^{out}\}=\sigma^2 \{\eta_{k,i}\} 
\end{align}

\subsection{Constructing the model}
The MPC model of Astlingen network can now be constructed considering the interconnection of the tanks and delays presented in Figure \ref{fig:1} and using the models discussed above. The inflow of each considered subpart of the network are summarized in Table \ref{tab:1}. The $i$-th tank and the delay flow to it are noted by $T_i$ and $\eta_{i:j}$ respectively, with $j$ being the remaining delay in minutes to the tank. The outflow of subpart $z$ is written as $q_{k,z}^{out}$, and the $i$-th run-off inflow to the system is given by $w_{k,i}$.

\begin{table}[]
\tiny
\begin{tabular}{lllll}\hline
 \textbf{Subpart}&  \textbf{Inflow}&  \textbf{Subpart}&  \textbf{Inflow}  \\ \hline
 $T_1$ & $q^{out}_{k,\eta_{1:5}}$ & $\eta_{1:5}$ & $q^{out}_{k,T_2}+q^{out}_{k,\eta_{1:10}}$ \\
 $T_2$ & $w_{k,2}$ & $\eta_{1:10}$ & $w_{k,1}+q^{out}_{k,T_2}+q^{out}_{k,T_4}+q^{out}_{k,\eta_{1:15}}$ \\
 $T_3$ & $w_{k,3} + q^{out}_{k,\eta_{3:5}}$ & $\eta_{1:15}$ & $q^{out}_{k,T_5}$ \\
 $T_4$ & $w_{k,4}$ & $\eta_{3:5}$ & $q^{out}_{k,\eta_{3:10}}$ \\
 $T_5$ & $w_{k,5}$ & $\eta_{3:10}$ & $q^{out}_{k,\eta_{3:15}}$ \\
 $T_6$ & $w_{k,6}$ & $\eta_{3:15}$ & $q^{out}_{k,T_6}$ \\ \hline
\end{tabular}
\caption{Inflows to the different elements of the systems}\label{tab:1}
\end{table}

\section{MPC design}
The design of controllers used in this work for both MPC and CC-MPC are based on the models discussed above and the minimization of a cost that considers the following operational objectives for the network:
\begin{itemize}
\item Maximizing flow to the WWTP
\item Minimizing flow to the river/creek 
\item Minimizing roughness of control
\end{itemize}
The first objective can be achieved by a linear negative cost on the outflow of tank 1, while the second objective can be formulated as a linear positive cost on the total overflow of the system; these objectives are collectively written as $\textbf{z}_k$, with the weight $\textbf{Q}$. The third objective can be written as a quadratic cost on the change in control flow $\Delta q^u_k$, with the diagonal weight $R$. Due to the overflow being modeled by a penalty approach, a fourth objective of minimizing the accumulated overflow volume $\textbf{V}_k^w$ is introduced, with the weight $\textbf{W}$.
\begin{align} 
\label{eq:43} J =  \min\limits_{\textbf{q}^u,\textbf{q}^w} \Sigma_{k=0}^N& \|\Delta\textbf{q}_k^u \|_R^2  + \textbf{Q}^T \textbf{z}_k   + \textbf{W}^T \textbf{V}_k^w \\
\text{subject to}&\nonumber\\
\label{eq:44} \textbf{z}=&\Phi_{Con}\textbf{q}^u  +\Psi\textbf{V}_0  +\Theta\textbf{w} +\Gamma\textbf{q}^w \\
\label{eq:45} \textbf{V}_k^w=& \Sigma_{i=0}^k\Delta T\textbf{q}_i^w
\end{align}

By using the MPC model over the prediction horizon $N$, the cost function of the MPC can be written as in (\ref{eq:43}), while the predicted objectives $\textbf{z}$ and accumulated overflow volumes, given by (\ref{eq:44}) and (\ref{eq:45}), are derived by substitution of the predicted volumes and delays. The constraints of the MPC model can similarly be collected into a single matrix inequality given by 
\begin{equation}
\label{eq:46} \Omega_{Con}\textbf{q}^u+\Omega_{vol} \textbf{V}_0+\Omega_{rain} \textbf{w}+\Omega_{weir}\textbf{q}^w\leq\pmb{\Omega}
\end{equation}
where the subscripts of the $\Omega$ matrix terms relates to the corresponding terms: $\textit{Con}$ for the control term, $\textit{vol}$ for the initial volume term, $\textit{rain}$ for the external inflows term, and $\textit{weir}$ for the term describing the CSOs of the system.

The design of the CC-MPC can similarly be derived using the corresponding model presented above. The cost of the resulting optimization program, appear as the expectation of (\ref{eq:43}) with the added linear cost term of the minimization of the slack variables \textbf{c} and \textbf{s} with weights $\textbf{W}_c$ and $\textbf{W}_s$
\begin{equation}
\label{eq:47} \begin{split}
J =  \min\limits_{\textbf{q}^u,\textbf{q}^w,\textbf{c},\textbf{s}}E\{\Sigma_{k=0}^N \|\Delta\textbf{q}_k^u\|_R^2  + \textbf{Q}^T \textbf{z}_k\\   + \textbf{W}^T \textbf{V}_k^w\} +\textbf{W}_c^T \textbf{c}+\textbf{W}_s^T \textbf{s}
\end{split}
\end{equation}
The expected objectives are given by 
\begin{equation}
\label{eq:48} E\{\textbf{z}\}=\Phi_{Con}\textbf{q}^u  +\Psi E\{\textbf{V}_0 \}  +\Theta E\{\textbf{w}\}  +\Gamma \textbf{q}^w
\end{equation}
while the accumulated overflow volume is unchanged from (\ref{eq:45}).

The matrix inequality of the collected probabilistic constraints are given by
\begin{equation}
\label{eq:49}\begin{split}
 \Omega_{Con} \textbf{q}^u+\Omega_{vol} E\{\textbf{V}_0\}+\Omega_{rain} E\{\textbf{w}\}+\Omega_{weir} \textbf{q}^w\leq\\ \pmb{\Omega}-\sigma\{\Omega_{vol}\textbf{V}_0+\Omega_{rain}\textbf{w}\} \Phi^{-1} (\gamma)+\Omega_s\textbf{s}+\Omega_c\textbf{c} 
\end{split}
\end{equation}
and the variance term
\begin{equation}
\label{eq:50} \sigma^2 \{\Omega_{vol} \textbf{V}_0+\Omega_{rain} \textbf{w}\}=\Xi_{vol} \sigma^2 \{\textbf{V}_0\}+\Xi_{rain} \sigma^2\{\textbf{w}\} 
\end{equation}
The weighting of the different objectives in the cost functions is done in accordance with the penalty approach\cite{HFC,JLS1}. The priority of the different objectives is given in the following order from highest to lowest priority:
\begin{enumerate}
\item Minimization of accumulated overflow volume $\textbf{V}_k^w$
\item Minimization of flow to the river/creek
\item Maximizing flow to the WWTP
\item Minimizing roughness of control
\end{enumerate}
The weightings used in this work are for the accumulated overflow volume given in Table \ref{tab:2} for each tank weir. The weights of the remaining objectives are 2 for the flow to the river/creek, -1 for the flow to the WWTP, 0.01 for the roughness of the control, and in the CC-MPC case 10 for the usage of the slack variables. The weights indicate that the avoidance of the flow to the river is prioritized twice as high as increasing flow to the WWTP. The weight on the roughness indicates the desire for the control to be smooth, but not a general priority. As seen from the table, the priority of the accumulated overflow is significantly higher than the other objectives.
\begin{table}[]
\centering
\begin{tabular}{llllll}\hline
\textbf{T1} & \textbf{T2} & \textbf{T3} & \textbf{T4} & \textbf{T5} & \textbf{T6} \\ \hline
1000 & 5000 & 5000 & 5000 & 5000 & 10000\\ \hline
\end{tabular}
\caption{Cost function weighting of accumulated overflow volume \textbf{W}, showing a higher cost for upstream elements.}\label{tab:2}
\end{table}

\section{Results}
The CC-MPC discussed above has been applied to the SWMM model of the Astlingen benchmark network. In order to test the strength of the CC-MPC, different types of uncertainty have been applied. Four different scenarios have been tested with the first being variations in the probability confidence level $\gamma$, changing from 60\% to 100\%. The remaining three scenarios are related with the uncertainty itself and how the MPC relies on its forecast information, where one scenario varies the size of the bound of the uncertainty and the last two varies the expected values of the inflow prediction’s deviation from the actual inflow, through scaled and offset biases. 
 During each test, only one parameter has been changed. In the baseline test case, the CC-MPC has been designed using a 90\% probability confidence level, a 50\% uncertainty bound, 0\% scaled bias and zero offset bias. In all the simulations, the uncertainty has been assumed that it follows a truncated normal distribution, where the lower bound is zero and the upper bound is three standard deviations above the expected disturbance.

\subsection{CC-MPC with Various Probability Confidence Levels $\gamma$}
\begin{figure}
\includegraphics[width=0.5\textwidth,trim={6cm 4.5cm 6cm 3cm},clip]{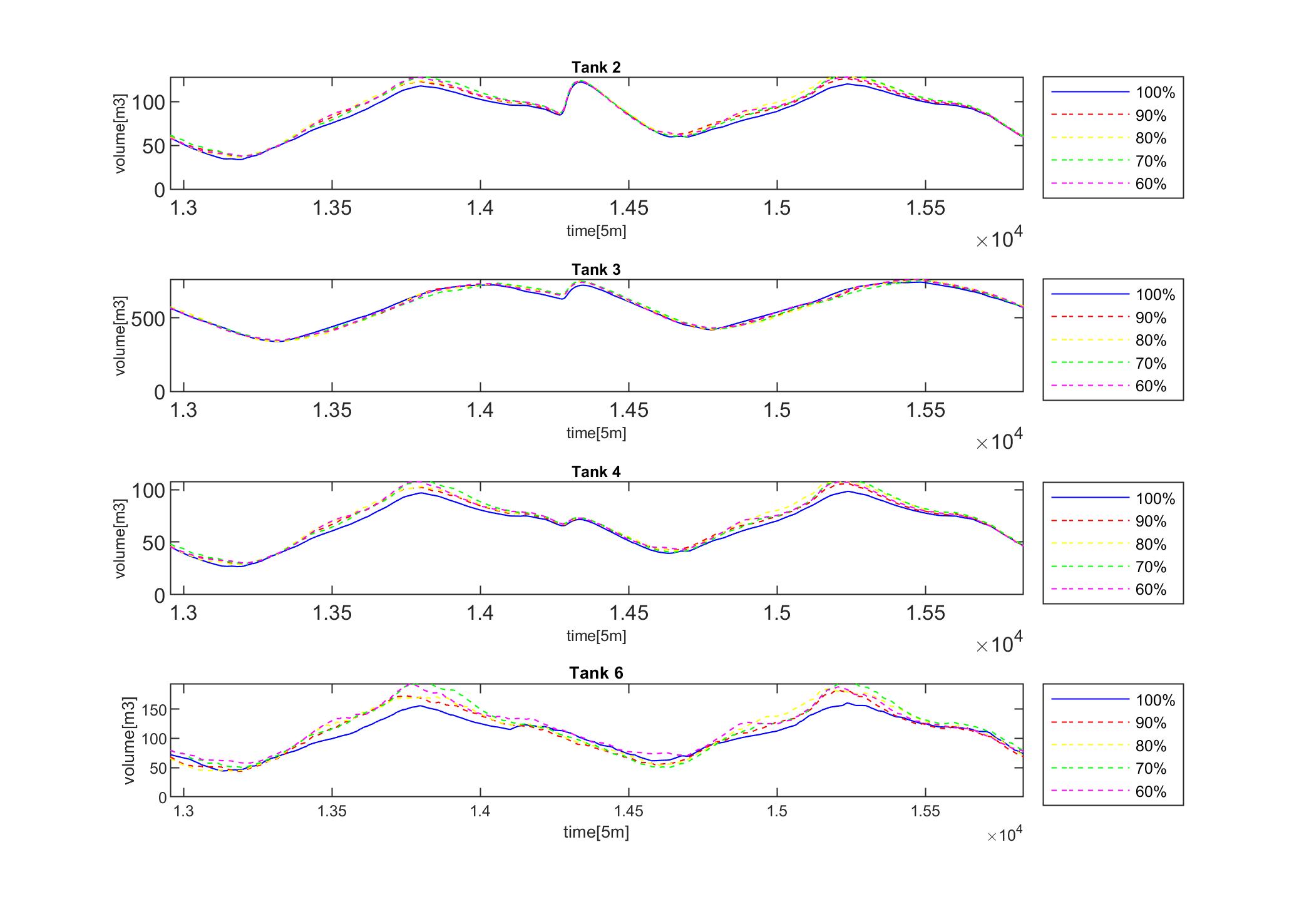}
\caption{The volumes for the tanks with controllable orifices (Tank 2, Tank 3, Tank 4, Tank 6) for the CC-MPCs with probability confidence levels $\gamma$ of 100-60\%.}\label{fig:2}
\end{figure}

The results in terms of CSO volume from varying the probability confidence level can be observed in Table \ref{tab:3}, and in Table \ref{tab:4} for the volume of treated water in WWTP. From these tables, we can see the distribution of CSO through the system. Both the CSO and WWTP volume of the CC-MPCs are comparatively close to the results of the deterministic MPC, regardless of the chosen probability guaranty. Similar conclusions can be obtained from Figure \ref{fig:2}, which presents volume dynamics for the tanks with controllable orifices (Tank 2, Tank 3, Tank 4, Tank 6) under CC-MPCs with probability confidence levels in the range from 60\% to 100\%. In Figure \ref{fig:2}, there are small deviations for the tank volumes resulting from CC-MPCs with different probability confidence levels. However, a slightly trend can be observed such that the smaller the probability confidence levels, the larger volumes at the peak points, which may reach the maximal storage more easily and generate more CSOs for the corresponding tanks. This figure only presents simulation results for day 10 and day 11 in order to provide a clearer view.
\begin{table*}[t]
\footnotesize
\centering
\begin{tabular}{lrrrrrr} \hline
\textbf{Tank \&}	&\textbf{MPC} &	\textbf{CC-MPC}&	\textbf{CC-MPC}&	\textbf{CC-MPC}&	\textbf{CC-MPC}&	\textbf{CC-MPC}\\
\textbf{Pipes}		&  &	\textbf{100\%}&\textbf{90\%}&\textbf{80\%}&\textbf{70\%}&\textbf{60\%}\\ \hline
T1	&93251	&93713	&92927	&93015&	93114&	93229\\
T2	&15484	&15683	&15544	&15543&	15543&	15543\\
T3	&34017	&34174	&34313	&34214&	34427&	34248\\
T4	&4814	&4823	&4814	&4814&	4814	&4815\\
T5	&15147	&15147	&15147	&15147&	15147&	15147\\
T6	&37950	&37723	&37946	&37939&	37980&	37870\\ \hline
P7	&4016	&4016	&4015	&4016&	4016	&4016\\
P8	&16207	&16207	&16191	&16203&	16203&	16199\\
P9	&4030	&4030	&4029	&4029&	4029	&4029\\
P10	&4838	&4838	&4842	&4839&	4839	&4840\\ \hline
River	&183754	&184585	&183778	&183774&	184086&	184020\\
Creek&	45996&	45769&	45990&	45984&	46025&	45915\\ \hline
Total	&229750	&230353	&229768	&229758	&230111	&229935\\ \hline
R. \%&		&-0.4522\%&	-0.0131\%&-0.0109\%&	-0.1807\%&	-0.1448\%\\
C. \%&		&0.4935\%&	0.0130\%	&0.0261\%&	-0.0630\%&	0.1761\%\\ \hline
Tot. \%&		&-0.2625\%&	-0.0078\%&-0.0035\%&	-0.1571\%&	-0.0805\%\\ \hline
\end{tabular}
\caption{Overflow results of the SWMM simulations with different controllers: MPC, and CC-MPC with the probability guarantees of 100-60\%}\label{tab:3}
\end{table*}
\begin{table*}[t]
\footnotesize
\centering
\begin{tabular}{p{1.5cm}rrrrrr} \hline
	&\textbf{MPC} &	\textbf{CC-MPC}&	\textbf{CC-MPC}&	\textbf{CC-MPC}&	\textbf{CC-MPC}&	\textbf{CC-MPC}\\
	&  &	\textbf{100\%}&\textbf{90\%}	&\textbf{80\%}&\textbf{70\%}&\textbf{60\%}\\ \hline
WWTP Vol.&	3772057	&3771560	&3772159&	3772088&	3771889&	3771795\\ \hline
Imp. \%	&	&-0.0132\%	&0.0027\%&	0.0008\%	&-0.0045\%&	-0.0069\%\\ \hline
\end{tabular}
\caption{Treated Wastewater results of the SWMM simulations with different controllers: MPC, and CC-MPC with the probability guarantees of 100-60\%}\label{tab:4}
\end{table*}

\subsection{CC-MPC with Various Uncertainty Bounds}
\begin{figure}
\includegraphics[width=0.5\textwidth,trim={4.5cm 3.5cm 5.5cm 3cm},clip]{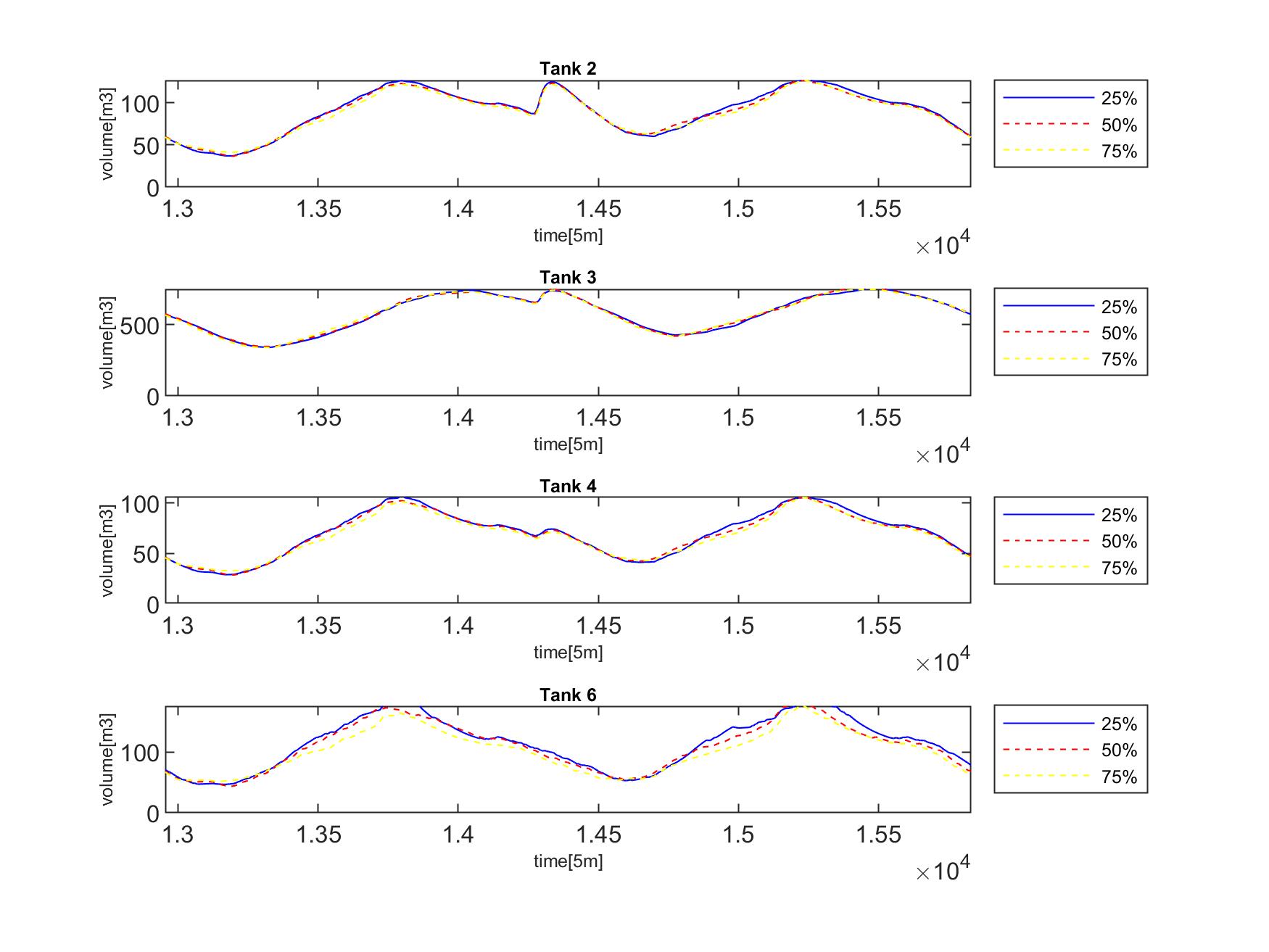}
\caption{The volumes for the tanks with controllable orifices (Tank 2, Tank 3, Tank 4, Tank 6) for the CC-MPC with the uncertainty bound of 25-75\%}\label{fig:3}
\end{figure}

The uncertainty bound describes the interval the uncertainty can take. For these simulations, a constant lower bound of zero is used; while the upper bound is defined as a percentage $p$ of the actual inflow above the expected rain inflow, see (\ref{eq:51}). The standard deviation of the uncertainty is assumed a third of the actual rain inflow times the percentage $p$, while the expectation is assumed equal to the actual rain. For normal distributions, this leads to the bound to be defined as
\begin{equation}
\label{eq:51} bound=[0,E\{q\}+p\mu]
\end{equation}
corresponding to the 99.7\% confidence interval of a corresponding unbounded distribution, if expectation matches the actual inflow.
The CC-MPC is tested with percentage $p$ bounds of 25\%, 50\% and 75\%. From Tables \ref{tab:5} and \ref{tab:6}, we can observe the resulting CSO volume and WWTP volume, respectively. It can be observed that the deviations from the results of the deterministic MPC are negligible of up to a few hundred cubic meters. Figure \ref{fig:3} provides detailed dynamic evolution for the tank volumes of CC-MPC with uncertainty bounds of 25\%, 50\% and 75\%, confirming conclusions obtained from Table \ref{tab:5} showing that the deviations brought by CC-MPCs are negligible. On another hand, it can be observed from Figure \ref{fig:3} that, the larger the uncertainty bound is, the smaller the tank volume is, which may cause less CSOs to the corresponding tank. This is because the larger uncertainty bounds make the CC-MPC generate more conservative orifice operations with the function of preventing CSOs. This conclusion is also in agreement with the basic deviations trends for the tanks CSO comparisons in Table \ref{tab:5}.
\begin{table}[]
\scriptsize
\begin{tabular}{lrrrr} \hline
\textbf{Tank \&} 	&\textbf{MPC}	&\textbf{CC-MPC}	&\textbf{CC-MPC}&	\textbf{CC-MPC}\\
\textbf{Pipes}	&	&\textbf{25\%}	&\textbf{50\%}&	\textbf{75\%}\\ \hline
T1	&93251	&93067&	92927&	92795\\
T2	&15484	&15543&	15544&	15544\\
T3	&34017	&34267&	34313&	34067\\
T4	&4814	&4814&	4814	&4814\\
T5	&15147	&15147&	15147&	15147\\
T6	&37950	&37939&	37946&	37673\\ \hline
P7	&4016	&4016&	4015	&4016\\
P8	&16207	&16203&	16191&	16207\\
P9	&4030	&4029&	4029	&4030\\
P10	&4838	&4839&	4842	&4838\\ \hline
River	&183754&	183879&	183778&	183412\\
Creek&	45996&	45984&	45990&	45718\\ \hline
Total	&229750&	229864	&229768	&229130\\ \hline
R. \%&	&	-0.0680\%	&-0.0131\%&	0.1861\%\\
C. \%&	&	0.0261\%	&0.0130\%&	0.6044\%\\ \hline
Tot. \%&	&	-0.0496\%	&-0.0078\%&	0.2699\%\\ \hline
\end{tabular}
\caption{Overflow results of the SWMM simulations with different controllers: MPC and CC-MPC with the uncertainty bound of 25-75\%.}\label{tab:5}
\end{table}
\begin{table}[]
\scriptsize
\begin{tabular}{p{1.5cm}lrrrr} \hline
 	&\textbf{MPC}	&\textbf{CC-MPC}	&\textbf{CC-MPC}&	\textbf{CC-MPC}\\
	&	&\textbf{25\%}	&\textbf{50\%}&	\textbf{75\%}\\ \hline
WWTP Vol.&	3772057	&3772086	&3772159	&3772676\\ \hline
Imp. \%	&	&0.0008\%&	0.0027\%	&0.0164\%\\ \hline
\end{tabular}
\caption{Treated Wastewater results of the SWMM simulations with different controllers: MPC, and CC-MPC with the uncertainty bound of 25-75\%.}\label{tab:6}
\end{table}

\subsection{CC-MPC with Various Scaled Biases}
\begin{figure}
\includegraphics[width=0.5\textwidth,trim={4.5cm 3.5cm 5.5cm 3cm},clip]{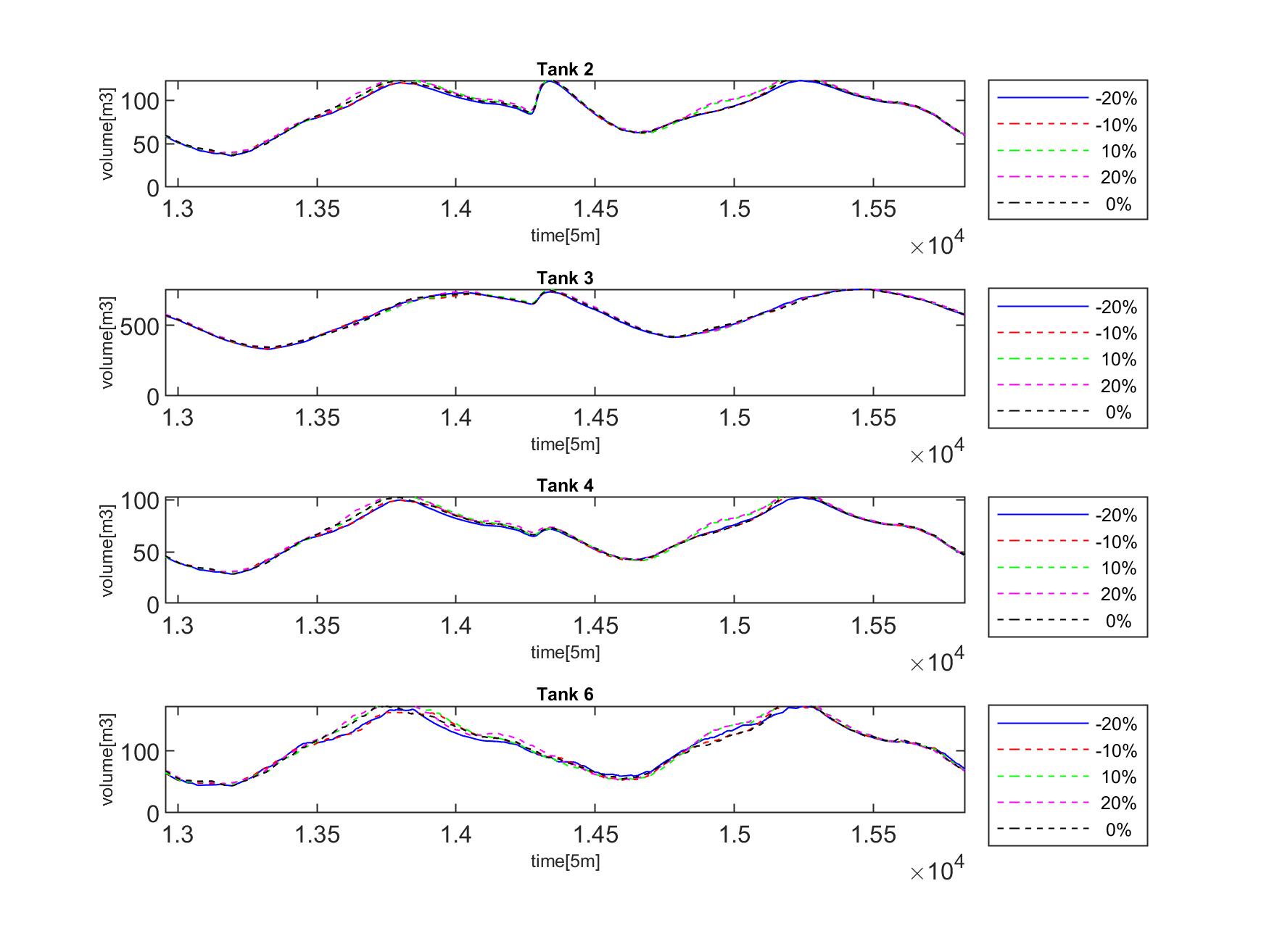}
\caption{The volumes for the controllable tanks under CC-MPC with different scaled bias.}\label{fig:4}
\end{figure}

In this section, the percentage bound on the uncertainty are kept constant, 50\%, instead the expected inflow is introduced as a scaled version of the actual rain inflow, given by 
\begin{equation}
\label{eq:52} E\{q\}=aq^{actual}
\end{equation}
Both the CC-MPC and the MPC are tested with 20\% and 10\% underestimated inflow, perfect forecast, and 10\% and 20\% overestimated inflow. The results can be seen in Table \ref{tab:7} and \ref{tab:8}, for the CSO volume and the WWTP volume, respectively. We can observe that if the expected inflow is overestimated then both types of MPC perform relatively worse as the overestimation increases with respect to CSO volume, and slight improvement of WWTP volume. When the inflow is underestimated, then the MPC performs significantly worse than the MPC with perfect forecast, when regarding CSO but only slightly better for the WWTP volume. For the CC-MPC, both the total CSO and WWTP results are relatively close to the MPC with perfect forecast, but with the drawback of the distribution of the CSOs being significantly worse for the creek. Figure \ref{fig:4} gives detail volume comparisons for the controllable tanks under CC-MPC with different scaled bias through a two-day simulation (day 10 and day 11). The dynamics of Figure \ref{fig:4} confirm that CC-MPC with an underestimated inflow performs significantly worse than that the CC-MPC with overestimated inflows. The explanation for this conclusion is also due to less conservative generated by the underestimated inflows. Moreover, the larger scales tend to have more differences in terms of tank volumes.
\begin{table*}[t]
\scriptsize
\centering
\begin{tabular}{p{1.5cm}|rr|rr|rr|rr|rr} \hline
\textbf{Tank \&}   &\textbf{ MPC}&\textbf{CC-MPC}&\textbf{MPC}&\textbf{CC-MPC}&\textbf{MPC}&\textbf{CC-MPC}&\textbf{MPC}&\textbf{CC-MPC}&\textbf{MPC}&	\textbf{CC-MPC}\\
\textbf{Pipes} &\textbf{ -20\%}& \textbf{ -20\%}& \textbf{-10\%}& \textbf{-10\%}& \textbf{0\%}& \textbf{0\%}& \textbf{10\%}	& \textbf{10\%}	& \textbf{20\%}	& \textbf{20\%} \\ \hline
T1	&96776&	90004&	95187	&91355	&93251&	92927&	94419&	94728&	96383	&96615 \\ 
T2	&16727&	16801&	15957	&16023	&15484&	15544&	15384&	15383&	15317	&15316 \\ 
T3	&33182&	33298&	33842	&33857	&34017&	34313&	34239&	34065&	33928	&34304 \\ 
T4&	5938	&5960	&5191	&5206	&4814&	4814&	4730&	4729&	4714	&4713 \\ 
T5	&15147	&15147&	15147	&15147	&15147	&15147&	15147&	15147&	15147&	15147 \\ 
T6	&39252	&39082&	38341	&38296	&37950	&37946&	37790&	37770&	37908&	37836 \\  \hline
P7	&4015	&4015&	4016	&4015	&4016&	4015&	4015	&4016&	4015	&4015 \\ 
P8	&16195	&16190&	16208	&16195	&16207	&16191&	16188&	16203&	16188&	16191 \\ 
P9	&4029	&4029&	4030	&4029	&4030&	4029&	4028&	4029&	4028	&4029 \\ 
P10	&4841	&4843&	4837	&4841	&4838&	4842&	4843&	4839&	4843	&4842 \\  \hline
River	&188805&	182242&	186369	&182623	&183754	&183778&	184949&	185094&	186519&	187129 \\ 
Creek&	47297&	47126&	46387&	46341	&45996	&45990	&45834&	45815&	45952&	45880 \\  \hline
Total	&236102&	229368&	232756&	228964&	229750&	229768&	230782&	230909&	232470	&233008 \\  \hline
R. \%&	-2.7488	&0.8228	&-1.4231&	0.6155&		&-0.0131&	-0.6503&	-0.7292&	-1.5047	&-1.8367 \\ 
C. \%&	-2.8285	&-2.4567	&-0.8501&	-0.7501&		&0.0130&	0.3522&	0.3935&	0.0957	&0.2522 \\  \hline
Tot. \%&	-2.7647	&0.1663	&-1.3084&	0.3421&		&-0.0078&	-0.4492&	-0.5045&	-1.1839	&-1.4181 \\  \hline
\end{tabular}
\caption{Overflow results of the SWMM simulations with different controllers: MPC and CC-MPC under different scaled bias.}
\label{tab:7}
\end{table*}
\begin{table*}[t]
\scriptsize
\centering
\begin{tabular}{p{1.5cm}|rr|rr|rr|rr|rr} \hline
   &\textbf{ MPC}&\textbf{CC-MPC}&\textbf{MPC}&\textbf{CC-MPC}&\textbf{MPC}&\textbf{CC-MPC}&\textbf{MPC}&\textbf{CC-MPC}&\textbf{MPC}&	\textbf{CC-MPC}\\
 &\textbf{ -20\%}& \textbf{ -20\%}& \textbf{-10\%}& \textbf{-10\%}& \textbf{0\%}& \textbf{0\%}& \textbf{10\%}	& \textbf{10\%}	& \textbf{20\%}	& \textbf{20\%} \\ \hline
WWTP Vol.&	3765554&	3772166&	3769365&	3772992&	3772057&	3772159&	3771015&	3770672&	3769214&	3768942\\ \hline
Imp. \%	&-0.1724	&0.0029	&-0.0714&	0.0248&		&0.0027&	-0.0276&	-0.0367&	-0.0754&	-0.0826\\ \hline
\end{tabular}
\caption{Treated Wastewater results of the SWMM simulations with different controllers: MPC, and CC-MPC under different scaled bias.}\label{tab:8}
\end{table*}

\subsection{CC-MPC with Various with Offset Biases}
\begin{figure}
\includegraphics[width=0.5\textwidth,trim={4.5cm 3.4cm 5.5cm 3.0cm},clip]{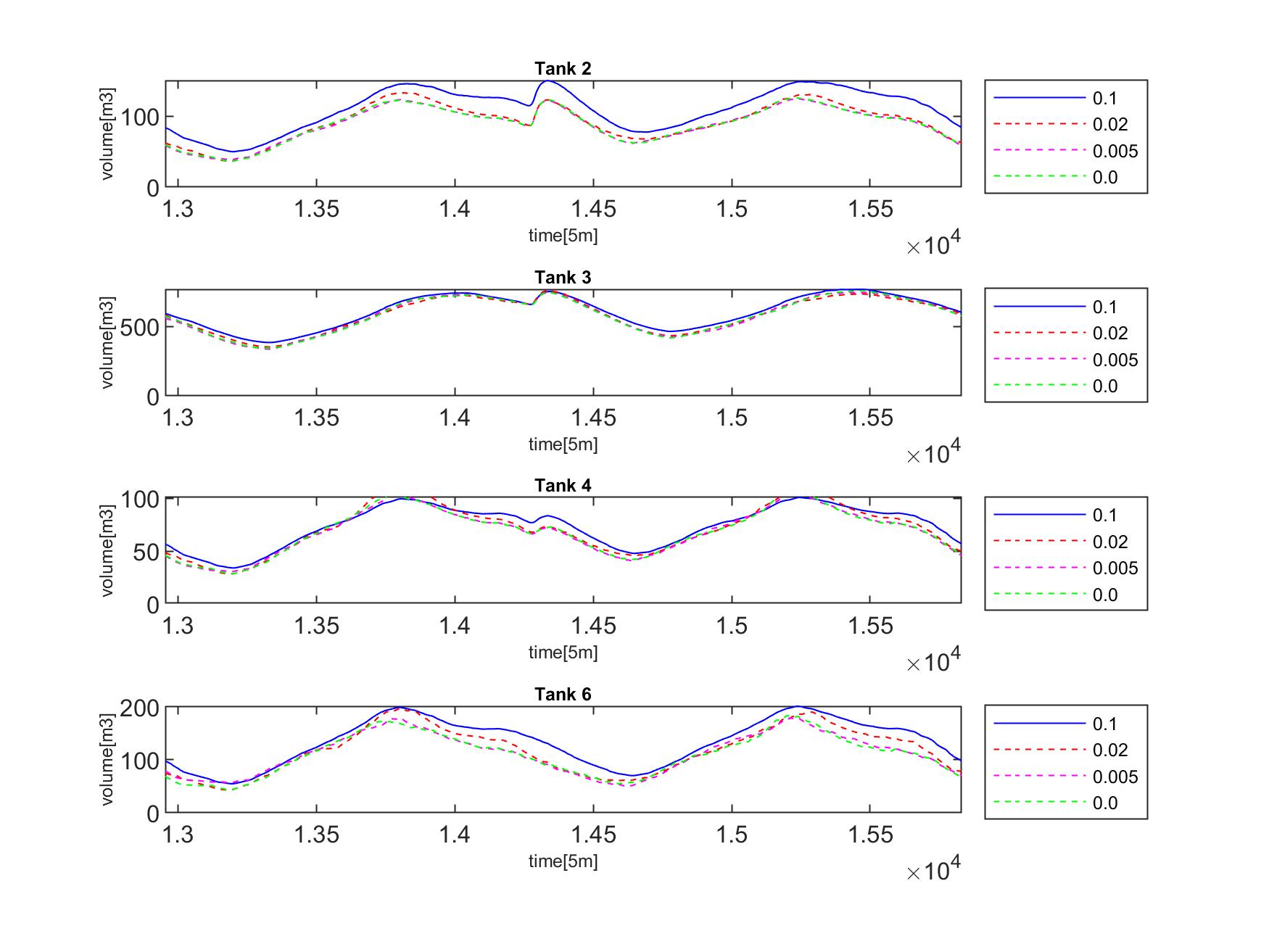}
\caption{The volumes for the controllable tanks under CC-MPC using different offsets.}\label{fig:5}
\end{figure}

In this section, the bias is changed from a scaling to an offset, see (\ref{eq:53}). Both the CC-MPC and the MPC are tested with zero offset and three positive offsets. The sizes of the offsets are the annual mean inflow (0.02) times the factors of 1 and 0.25, and 10 times the dry-weather inflow (0.1)
\begin{equation}
\label{eq:53} E\{q\}=q^{actual}+b
\end{equation}
The results of both MPC types can be seen in Table \ref{tab:9} and \ref{tab:10} for the CSO and WWTP volume, respectively. We can observe that for both non-zero offsets, the CSO is significantly worse, with the offset of 0.1 being even worse. The results of the WWTP volume are also worse than the MPC with perfect forecast. Figure \ref{fig:5} gives more information about the performance of CC-MPC under different offsets. The differences in tank volume among CC-MPC using different offsets are compared. As always, the more volume in the tank indicates an increased chance of having more CSOs. From Figure \ref{fig:5}, we can conclude that CC-MPC with 0.1 offset have more tank volume than that the offsets, which means, CC-MPC with 0.1 offset behaves worse than that of MPC. However, the CC-MPC with 0.005 and 0.02 did not show a clear trend.
\begin{table*}[t]
\small
\begin{tabular}{p{1.5cm}|rr|rr|rr|rr} \hline
\textbf{Tank \&}   &  \textbf{MPC} &  \textbf{CC-MPC} &  \textbf{MPC} &  \textbf{CC-MPC} & \textbf{MPC}  &  \textbf{CC-MPC} & \textbf{MPC} & \textbf{CC-MPC} \\
\textbf{Pipes} &   &  \textbf{0} &  \textbf{0.005}&  \textbf{0.005} & \textbf{0.02}  & \textbf{0.02} & \textbf{0.1}& \textbf{0.1}\\  \hline
T1&	93251&	92927&	93655&	93856&	96472&	96590&	131407&	130211\\
T2&	15484&	15544&	15387&	15450&	15453&	15452&	15847&	15511\\
T3&	34017&	34313&	33975&	34322&	34086&	34485&	36811&	36548\\
T4&	4814	&4814	&4728	&4728	&4639	&4644	&4465	&4465\\
T5&	15147&	15147&	15147&	15147&	15147&	15147&	15147&	15147\\
T6&	37950&	37946&	37916&	37961&	37877&	37780&	37907&	37763\\ \hline
P7&	4016	&4015	&4015	&4015	&4015	&4016	&4016	&4016\\
P8&	16207&	16191&	16188&	16193&	16188&	16203&	16203&	16203\\
P9&	4030	&4029	&4028	&4029	&4028	&4029	&4029	&4029\\
P10&	4838	&4842	&4843	&4842	&4843	&4839	&4839	&4839\\ \hline
River&	183754	&183778	&183922	&184536	&186828	&187360	&224718&	222925\\
Creek&	45996	&45990	&45959	&46005	&45920	&45825	&45952&	45808\\ \hline
Total&	229750	&229768	&229881	&230541	&232748	&233185	&270670&	268733\\  \hline
R. \%&	&	-0.0131&	-0.0914&	-0.4256&	-1.6729&	-1.9624&	-22.2928&	-21.3171\\
C. \%&	&	0.0130&	0.0804&	-0.0196&	0.1652&	0.3718&	0.0957&	0.4087\\  \hline
Tot. \%&	&	-0.0078&	-0.0570&	-0.3443&	-1.3049&	-1.4951&	-17.8107&	-16.9676\\  \hline
\end{tabular}
\caption{Overflow results of the SWMM simulations with different controllers: MPC and CC-MPC under different off-set biases.}\label{tab:9}
\end{table*}
\begin{table*}[t]
\small
\begin{tabular}{p{1.5cm}|rr|rr|rr|rr}  \hline
 &  \textbf{MPC} &  \textbf{CC-MPC} &  \textbf{MPC} &  \textbf{CC-MPC} & \textbf{MPC}  &  \textbf{CC-MPC} & \textbf{MPC} & \textbf{CC-MPC} \\
 &   &  \textbf{0} &  \textbf{0.005}&  \textbf{0.005} & \textbf{0.02}  & \textbf{0.02} & \textbf{0.1}& \textbf{0.1}\\  \hline
WWTP Vol. & 3772057 & 3772159 & 3771978 & 3771575 & 3768823 &3768643 &3731689 &3733651 \\  \hline
Imp. \% &  & 0.0027 &-0.0021  &0.0001  &-0.0857  & -0.0905 & -1.0702& -1.0182 \\  \hline
\end{tabular}
\caption{Treated Wastewater results of the SWMM simulations with different controllers: MPC, and CC-MPC under different off-set biases.}\label{tab:10}
\end{table*}

From the above results, we can infer that the CC-MPC is capable of handling different type of uncertainties, and for those type of uncertainties, it performs similarly to the deterministic MPC. We can further see that the CC-MPC, while not performing that well with constant offset biases, these biases were also outside the uncertainty bound, practically making the CC-MPC as blind as the deterministic MPC. In real-world scenarios, the uncertainty of the inflow is not exactly as the one used here. Instead the uncertainty bound would vary across the prediction horizon, as would do the biases of the expected inflow.

\section{Conclusion}
A stochastic MPC has been applied to a hydrodynamic SWMM model of the Astlingen urban drainage benchmark network, using a chance-constraint formulation of MPC. A comparison study of the application of both CC-MPC and MPC has been done for several scenarios and types of uncertainties in forecasts, involving both biases in the forecast to different sizes of the uncertainty. Based on the simulations, we can conclude that only the uncertainty regarding biases has an effect on the performance of CC-MPC. Furthermore, it could be observed that the performance of both type of MPC considered deteriorate similarly with respect to CSO volume, when the forecast overestimates the rain inflow. However, when the forecast underestimates the rain inflow, then the CC-MPC performs similarly to the ideal case, while the performances of deterministic MPC deteriorates.



\bibliographystyle{elsarticle-num}
\bibliography{SWMM_CC-MPC_Manuscript_CEP}


\end{document}